# Thermoelectric Effect at Quantum Limit in Two-Dimensional Organic Dirac Fermion System with Zeeman Splitting

Toshihito Osada*

*Institute for Solid State Physics, University of Tokyo,*

*5-1-5 Kashiwanoha, Kashiwa, Chiba 277-8581, Japan.*

The thermoelectric effect in a two-dimensional (2D) massless Dirac fermion (DF) system at the quantum limit is discussed to verify the prediction of high-performance thermopower in an organic conductor $\alpha$-(BEDT-TTF)$_2$I$_3$. Because of relatively large Zeeman splitting in $\alpha$-(BEDT-TTF)$_2$I$_3$, the boundless increase of thermopower at high magnetic fields, predicted without the Zeeman effect, is hardly expected, whereas there appears to be a broad local maximum. This is characteristic of 2D DF systems with Zeeman splitting and is recognized in the previous experiment. In contrast to 3D Dirac/Weyl semimetals with robust gapless features, it might be difficult to realize high-performance thermopower in real 2D DF systems under high magnetic fields.

The thermoelectric effects of topological semimetals have recently attracted a great deal of attention. For three-dimensional (3D) Dirac or Weyl semimetals at the high-magnetic-field quantum limit, it has been shown that the thermoelectric Hall conductivity $\alpha_{xy}^{(3D)}$ takes a constant value that is proportional to temperature but independent of the magnetic field and/or carrier density, which is owing to the constant density of states of the $N=0$ chiral Landau subband [1,2]. Correspondingly, the Seebeck coefficient $S_{xx}$ increases linearly with no saturation as the magnetic field increases, resulting in high-performance thermopower. This feature has been observed in a 3D Dirac semimetal $Pb_{1-x}Sn_xSe$ with a small spin-orbit gap [3] and was recently confirmed in a 3D Dirac semimetal $ZrTe_5$ [4].

In 2019, Liang Fu predicted a similar high-performance thermopower in two-dimensional (2D) Dirac fermion (DF) systems at the quantum limit and cited a layered organic conductor $\alpha$-$(BEDT-TTF)_2I_3$ as a candidate material [5]. The thermoelectric characteristics of 2D DFs under magnetic fields have already been investigated theoretically [6] to explain graphene experiments [7]. Fu particularly focused on the DF system with a small carrier density near charge neutrality. At the quantum limit, the Fermi level exists at the $N=0$ Landau level (LL) at zero energy. It has four-fold spin and valley degeneracy when the interaction and Zeeman effect can be neglected. In this situation, the 2D thermoelectric Hall conductivity $\alpha_{xy}$ acquires a universal quantized value $4(\log 2)k_B e/h$, which is independent of temperature as long as the LL width $\Gamma$ is much smaller than $k_B T$ (the dissipationless limit). This feature leads to unprecedented growth in the thermopower $S_{xx}$ and thermoelectric figure of merit under high magnetic fields even at low temperatures, where no realistic thermoelectric material is known, similar to the 3D Dirac/Weyl semimetals [5]. The thermoelectric features were also discussed for another 2D Dirac-like system, where a pair of Dirac cones merged [8].

In this study, we consider how the thermoelectric features predicted by Fu appear in real $\alpha$-(BEDT-TTF)$_2$I$_3$ at the dissipationless limit $\Gamma \ll k_\text{B}T$. $\alpha$-(BEDT-TTF)$_2$I$_3$ is usually regarded as a 2D system owing to its weak interlayer coupling. A 2D massless DF system with a pair of tilted Dirac cones (valleys) is realized in the high-pressure (>1.3 GPa) metallic phase [9]. This fact was originally found theoretically [10] and later experimentally confirmed using interlayer magnetotransport [11-13], specific heat [14], and nuclear magnetic resonance (NMR) [15].

The thermoelectric effect of the DF state in $\alpha$-(BEDT-TTF)$_2$I$_3$ under magnetic fields was experimentally investigated by Konoike *et al.* [16]. They found that the Seebeck coefficient (thermopower) $S_{xx}$ and, particularly, the Nernst coefficient $S_{xy}$ show a large local maximum in the temperature dependence under a fixed magnetic field. In addition, $S_{xx}$ and $S_{xy}$ exhibit an exponential decrease at low temperatures. These features were well reproduced by Proskurin *et al.* using quantum transport calculations considering scattering [17]. The local maximum corresponds to the thermal distribution to the $N = \pm 1$ LL, and the exponential decrease originates from the Zeeman splitting of the $N = 0$ LL. Apparently, we cannot find any sign of Fu's prediction in the experimental data.

Therefore, we constructed the same level of argument as Fu's for a more realistic $\alpha$-(BEDT-TTF)$_2$I$_3$ model and investigated the expected thermoelectric effect. Specifically, we considered the relatively large Zeeman splitting in $\alpha$-(BEDT-TTF)$_2$I$_3$ and investigated the dissipationless thermoelectric effect using an edge picture, which was first used by Girvin and Jonson (G-J) to discuss the thermoelectric effect in the quantum Hall state of a 2D electron gas at $\Gamma \ll k_B T$ [18].

The LLs of 2D massless DFs with Zeeman splitting are given by

$$\varepsilon_{N,\sigma} = \text{sgn}(N)\sqrt{2e\hbar v_F^2 |B_z||N|} + \sigma \mu_B |B_z|, \tag{1}$$

where $N$ is the LL index ($N = 0, \pm 1, \pm 2, \cdots$), and $\sigma = -1, +1$ (abbreviated as $\uparrow, \downarrow$) indicates electron spin. $v_F$ and $\mu_B$ are the constant velocity of DFs and Bohr magneton, respectively. We assumed that the g-factor was two. Each LL has two-fold valley degeneracy.

The magnetic field dependence of LLs is shown in Fig. 1(a), where $c$ is the scale length. We chose $c = 1.75$ nm and $v_F = 2.4 \times 10^4$ m/s for $\alpha$-(BEDT-TTF)$_2$I$_3$. In this case, $\mu_B / ecv_F = 1.38$ is satisfied, and the unit magnetic field and unit energy are given by $\hbar/ec^2 = 215$ T and $\hbar v_F/c = 9.03$ meV ~ 105 K, respectively. Compared to graphene with a larger $v_F \sim 10^6$ m/s, the Zeeman splitting is relatively large.

Following the G-J theory and Fu's argument, we consider a ribbon-shaped DF system with the left and right edges shown in Fig. 1(b). A schematic of the energy spectrum is illustrated in Fig. 1(c) with no Zeeman splitting for simplicity. In the bulk region, the spectrum is a set of LLs $\varepsilon_{N,\sigma}$ with no dependence on the center coordinates $X_0$ of the cyclotron motion. Around the left and right edges, the LL spectra depend on $X_0$ and show valley splitting, as indexed by $\pm$, forming the edge states $\varepsilon_{N,\sigma}^{\pm}(X_0)$. Note that the N=0 LL shows valley splitting to electron-like $\varepsilon_{0,\sigma}^+$ and hole-like $\varepsilon_{0,\sigma}^-$. When a small temperature difference $T_R - T_L$ is applied between two edges, the distribution function $f^0(E) = 1/[1 + \exp\{(E - \mu)/k_B T\}]$ takes different values $f_L^0(E)$ and $f_R^0(E)$ at the left and right edges, respectively. This causes a difference in edge current between the left and right edges. Each electron-like edge state, $\varepsilon_{N>0,\sigma}^{\pm}(X_0)$ or $\varepsilon_{N=0,\sigma}^{+}(X_0)$, carries the following net current along the left and right edges:

$$
\begin{aligned}
I_{N,\sigma}^{\pm} &= \frac{1}{2\pi l^2} \int_{-\infty}^{\infty} (-e)\left(-\frac{l^2}{\hbar}\frac{\partial \varepsilon_{N,\sigma}^{\pm}(X_0)}{\partial X_0}\right) f^0(\varepsilon_{N,\sigma}^{\pm}(X_0)) dX_0 \\
&= \frac{e}{h}\int_{\varepsilon_{N,\sigma}}^{\infty} \{-f_L^0(E) + f_R^0(E)\} dE = \frac{e}{h}\left\{\int_{\varepsilon_{N,\sigma}}^{\infty}\frac{E-\mu}{T}\left(-\frac{\partial f^0}{\partial E}\right) dE\right\}(T_R - T_L) \\
&= \frac{k_B e}{h}\left[-f^0(\varepsilon_{N,\sigma})\log f^0(\varepsilon_{N,\sigma}) - \{1 - f^0(\varepsilon_{N,\sigma})\}\log\{1 - f^0(\varepsilon_{N,\sigma})\}\right](T_R - T_L)
\end{aligned} \quad (2)
$$

Here, $l = \sqrt{\hbar/e|B_z|}$ is the magnetic length. The current carried by the hole-like edge state, $\varepsilon_{N<0,\sigma}^{\pm}(X_0)$ or $\varepsilon_{N=0,\sigma}^{-}(X_0)$, has the same final form. The current density response against the electric field and temperature gradient is expressed as $\mathbf{j} = \vec{\sigma}\mathbf{E} + \vec{\alpha}(-\nabla T)$, where $\vec{\sigma}$ and $\vec{\alpha}$ are the electric conductivity and thermoelectric conductivity, respectively. Therefore, the thermoelectric Hall conductivity of the 2D DF system at $\Gamma \ll k_B T$ is given by

$$
\alpha_{xy} = 2\frac{k_B e}{h}\sum_{N,\sigma}\left[-f^0(\varepsilon_{N,\sigma})\log f^0(\varepsilon_{N,\sigma}) - \{1 - f^0(\varepsilon_{N,\sigma})\}\log\{1 - f^0(\varepsilon_{N,\sigma})\}\right]. \quad (3)
$$

Here, factor 2 comes from valley degeneracy. The chemical potential $\mu$ included in $f^0(\varepsilon_{N,\sigma})$ is determined by the following condition for the imbalance between the 2D electron density ($n$) and 2D hole density ($p$):

$$
n - p = \frac{1}{2\pi l^2}\sum_{\sigma}\left[f^0(\varepsilon_{N=0,\sigma}) + 2\sum_{N>0} f^0(\varepsilon_{N,\sigma}) - \{1 - f^0(\varepsilon_{N=0,\sigma})\} - 2\sum_{N<0}\{1 - f^0(\varepsilon_{N,\sigma})\}\right]. \quad (4)
$$

Because the thermoelectric power tensor $\ddot{S}$ is defined by $\mathbf{E} = \ddot{\sigma}^{-1}\mathbf{j} + \ddot{S}\nabla T$, $\ddot{S} = \ddot{\sigma}^{-1}\ddot{\alpha}$. At the dissipationless limit, the diagonal elements of $\ddot{\sigma}$ and $\ddot{\alpha}$ are negligibly small, and the Hall conductivity is given by $\sigma_{xy} = -e(n-p)/B_z$. Therefore, the Seebeck coefficient $S_{xx}$ is expressed as

$$S_{xx} = \frac{\sigma_{xx}\alpha_{xx} - \sigma_{xy}\alpha_{yx}}{\sigma_{xx}\sigma_{yy} - \sigma_{xy}\sigma_{yx}} \simeq \frac{\alpha_{xy}}{\sigma_{xy}} = -\frac{\alpha_{xy}B_z}{e(n-p)}. \tag{5}$$

The magnetic field dependence of the chemical potential, $\mu$, is shown in Fig. 1(a) for several temperatures. Here, we assumed a small carrier imbalance $(n-p)c^2 = 3\times 10^{-4}$, which corresponds to the 2D carrier density $n-p = 10^{10}$ cm$^{-2}$ for $c = 1.75$ nm. At the quantum limit $|B_z|/(\hbar/ec^2) > 0.001$, $\mu$ lies almost on the $\varepsilon_{0,\uparrow}$ LL at low temperatures but deviates from the $\varepsilon_{0,\uparrow}$ LL to zero energy at high temperatures.

Figure 2(a) shows the 2D carrier density dependence of the 2D thermoelectric Hall conductivity $\alpha_{xy}$ for several temperatures at a fixed magnetic field, which is indicated by an orange vertical line in Fig. 1(a). Zeeman splitting is considered in the solid curves, whereas no splitting is assumed in the dashed curves. At sufficiently low temperatures, $\alpha_{xy}$ is a set of semi-elliptic peaks, whose local maxima correspond to half-filling of $\varepsilon_{N,\sigma}$ LLs, and they have a common universal height of $\alpha_{xy} = 2(\log 2)k_B e/h$. Around the charge neutrality point $(n-p)c^2 = 0$, $\alpha_{xy}$ becomes close to zero reflecting the Zeeman gap at zero energy.

On the other hand, if we neglect the Zeeman splitting, the low temperature $\alpha_{xy}$ would behave like the dashed curves, which is a set of peaks with double width and height. Here, the charge neutral point must become a local maximum with a height of $\alpha_{xy} = 4(\log 2)k_B e/h$. This is the quantized thermoelectric Hall conductivity, which was discussed by Fu, and the non-vanishing value of $\alpha_{xy}$ causes a high-performance thermoelectric effect at low temperatures [5].

Figure 2(b) shows the temperature dependence of $\alpha_{xy}$ at a fixed carrier density around the charge neutrality point (indicated by the vertical orange line in Fig. 2(a)) for several magnetic fields. The solid (dashed) curves indicate the case with (without) Zeeman splitting. In the zero splitting case (dashed curves), with decreasing temperature, $\alpha_{xy}$ decreases and reaches the plateau of the quantized thermoelectric Hall conductivity $4(\log 2)k_B e/h$. When Zeeman splitting is finite (solid curves), $\alpha_{xy}$ first approaches the no-splitting value $4(\log 2)k_B e/h$, then it decreases again, forming a shoulder-like structure, and finally reaches a magnetic-field-dependent constant value, forming a plateau.

Figure 3(a) illustrates the field dependence of LLs and the chemical potential $\mu$ for a small carrier density in a field range wider than that in Fig. 1(a). The yellow shaded region ($10^{-3} < |B_z|/(\hbar/ec^2) < 2.5 \times 10^{-1}$) corresponds to the quantum limit. In this region, $\mu$ remains at the $\varepsilon_{0,\downarrow}$ LL at low temperatures, whereas it approaches zero at high temperatures. In the higher field region, the LLs with the Zeeman splitting periodically cross the zero energy because the Zeeman part becomes dominant in the LL energy (1). This causes the periodic oscillation of $\mu$.

The corresponding field dependence of the Seebeck coefficient $S_{xx}$ is illustrated for the case with (solid curves) and without (dashed curves) Zeeman splitting in Fig. 3(b).

The yellow shaded region corresponds to the quantum limit. As the magnetic field increases, $S_{xx}$ exhibits quantum oscillations at low temperatures before the quantum limit. In the quantum limit region, $S_{xx}$ asymptotically approaches the dashed line $S_{xx} = -\{4(\log 2)k_B / h(n-p)\}B_z$ with no splitting. As Fu pointed out, the thermopower $S_{xx}$ increases boundlessly in proportion to the magnetic field. With the Zeeman splitting, $S_{xx}$ first tends to follow the dashed line, then deviates from it, decreases forming a hump structure with a local maximum, and finally converges to a curve for the low-temperature limit. At higher fields, $S_{xx}$ oscillates, reflecting the periodic oscillation of $\mu$.

Although $\alpha$-(BEDT-TTF)$_2$I$_3$ was expected to be one of the candidate materials of Fu's proposal, the relatively large Zeeman splitting breaks the quantization of $\alpha_{xy}$, and the boundless growth in $S_{xx}$ under high magnetic fields cannot be expected even at the dissipationless limit. Instead, the shoulder-like structure in the temperature dependence of $\alpha_{xy}$ and a hump-like local maximum in the field dependence of $S_{xx}$ emerge as remnants of the effects proposed by Fu. They are characteristic of the 2D DF system with Zeeman splitting, and never appear in 2D non-Dirac systems. These features are recognized in the $\alpha$-(BEDT-TTF)$_2$I$_3$ experimental data (in the region of $B_z > 2$ T and $T < 5$ K) [16].

In the 2D dissipationless systems, $\alpha_{xy}$ is determined only by the bulk LL configuration, as is seen in Eq. (3). Therefore, the above arguments are also applicable for the 2D DF systems with tilting and/or anisotropic Dirac cones like real $\alpha$-(BEDT-TTF)$_2$I$_3$, because their LL configuration is given by the same form as Eq. (1).

Finally, it should be noted that the present results are obtained for the dissipationless limit $k_B T \gg \Gamma$, which was assumed in the 3D Dirac/Weyl semimetals argument. Therefore, the present model cannot provide any results on the zero-temperature limit and dissipative scattering effects. At zero temperature, $\alpha_{xy}$ and $S_{xx}$ must be zero. In fact, theory that considers scattering concluded that $\alpha_{xy}$ decreases to zero with decreasing temperature [6]. For the dissipative effect, the Nernst coefficient $S_{xy}$, which we ignored because of the assumption of $\sigma_{xx} = 0$ and $\alpha_{xx} = 0$, becomes finite in a disordered system [6,17,19]. The observed peak in the temperature dependence of $S_{xx}$ and $S_{xy}$, which originates from the thermal distribution to $\varepsilon_{1,\uparrow}$, is reproduced by considering the disorder [17].

In conclusion, we investigated the thermoelectric effect at the high-magnetic-field quantum limit in a 2D massless DF system to verify the high-performance thermopower proposed by Fu in an organic DF system $\alpha$-(BEDT-TTF)$_2$I$_3$. We assumed the dissipationless limit as in the 3D topological semimetal arguments. The boundless increase in thermopower, predicted without the Zeeman effect, is hardly expected because of the relatively large Zeeman splitting in $\alpha$-(BEDT-TTF)$_2$I$_3$, whereas the local maximum structure characteristic of the DF system appears in the field dependence of the thermopower. This feature is recognized in the previous experimental data. In contrast to 3D Dirac/Weyl semimetals with robust gapless features, it might be difficult to realize high-performance thermopower in real 2D DF systems under high magnetic fields.

**Acknowledgements**

The author deeply thanks Professor Liang Fu (MIT) for arousing our interest in the thermoelectric effect of topological semimetals, and Dr. Takako Konoike (NIMS) for valuable discussions. He also thanks Dr. A. Kiswandhi, Dr. M. Sato, and Mr. T. Ochi (ISSP) for their daily discussions. This work was supported by JSPS KAKENHI Grant Numbers JP20H01860 and JP21K18594.

# References

*osada@issp.u-tokyo.ac.jp

**Figure 1** (Osada)

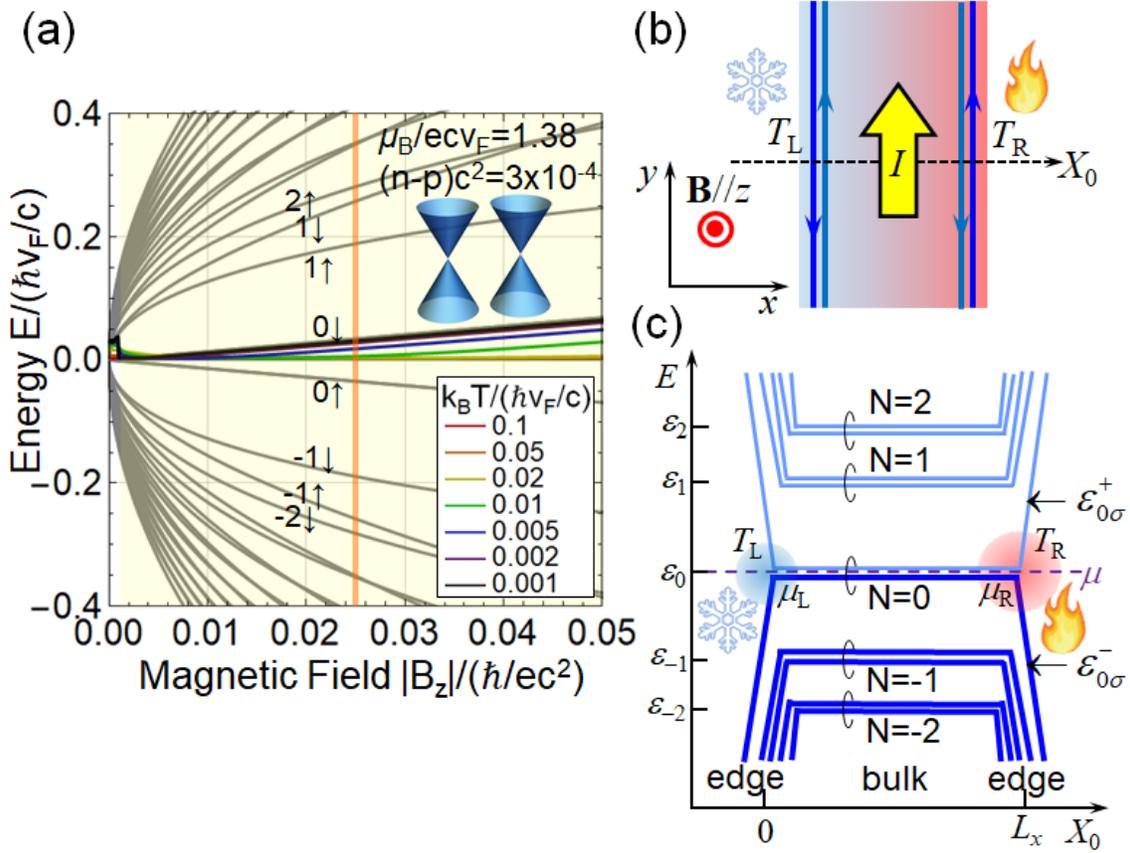

**FIG. 1.** (color online)

(a) Magnetic field dependence of the LLs and chemical potential $\mu$ for a fixed small carrier density at several temperatures in the 2D DF system. Inset shows the schematic of two valleys with Dirac-cone dispersion. (b) The configuration of the G-J edge picture for the thermoelectric effect under a magnetic field. (c) Schematic of the energy spectrum as a function of the center coordinate $X_0$ of the cyclotron motion of the 2D DF under a magnetic field. $L_x$ denotes the width of the ribbon.

**Figure 2** (Osada)

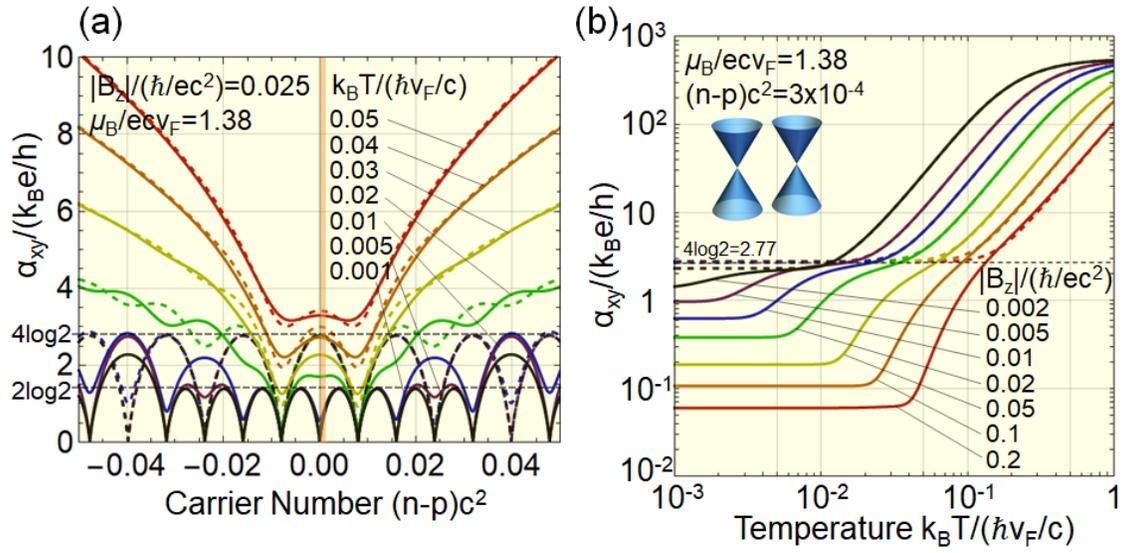

**FIG. 2.** (color online)

(a) 2D carrier density dependence of the 2D thermoelectric Hall conductivity $\alpha_{xy}$ for several temperatures at a fixed magnetic field. The unit of $\alpha_{xy}$ is $k_B e/h = 3.34$ nA/K. (b) The temperature dependence of $\alpha_{xy}$ for several magnetic fields at a fixed carrier density around the charge neutrality point. The solid and dashed curves are for cases with and without the Zeeman effect.

**Figure 3** (Osada)

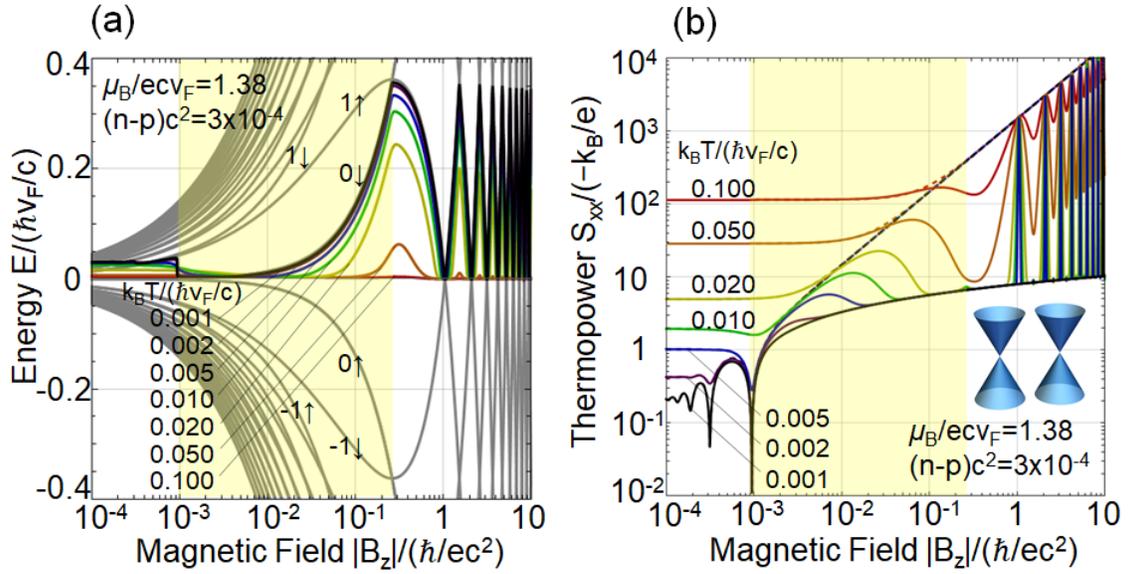

**FIG. 3.** (color online)

(a) The wide-range magnetic field dependence of the LLs and chemical potential $\mu$ for a fixed small carrier density at several temperatures in the 2D DF system. (b) The magnetic field dependence of the Seebeck coefficient $S_{xx}$ for several temperatures at a fixed carrier density around the charge neutrality point. The unit of $-S_{xx}$ is $k_B/e = 86.2\ \mu\text{V/K}$. The solid and dashed curves are for cases with and without the Zeeman effect.